\documentclass[conference]{IEEEtran}

\usepackage{ucs}
\usepackage[utf8x]{inputenc}
\usepackage[T1]{fontenc}
\usepackage[greek,english,ngerman]{babel}
\usepackage{bbold}  
\usepackage{url}

\usepackage{cite}
\usepackage{amsmath,amssymb,amsfonts}
\usepackage{graphicx}
\usepackage{textcomp}
\usepackage{xcolor}
\def\BibTeX{{\rm B\kern-.05em{\sc i\kern-.025em b}\kern-.08em
    T\kern-.1667em\lower.7ex\hbox{E}\kern-.125emX}}

\begin{document}
\selectlanguage{english}

\newcounter{definition}
\newcommand{\definition}{\refstepcounter{definition} {\noindent \bf Definition \thedefinition: }}
\renewcommand{\thedefinition}{\arabic{section}.\arabic{definition}}

\title{A reference model for interaction semantics}

\author{\IEEEauthorblockN{1\textsuperscript{st} Johannes Reich}
\IEEEauthorblockA{\textit{SAP SE} \\
69190 Walldorf, Germany\\
johannes.reich@sap.com}\\
\and
\IEEEauthorblockN{2\textsuperscript{nd} Tizian Schröder}
\IEEEauthorblockA{\textit{Otto-von-Guericke-University} \\
39106 Magdeburg, Germany\\
tizian.schroeder@ovgu.de}
}

\maketitle

\begin{abstract}
In this article, we introduce a reference model for interaction semantics among communicating discrete systems to guide the discourse on interoperability. 

The necessary set of unifying concepts is small and comprises essentially the notion of discrete systems interacting by exchanging information. It is based on a simple, but nevertheless complete classification of system interactions with respect to information transport and processing. Information transport can only be uni- or bidirectional and information processing is subclassified along the binary dimensions of state, determinism and synchronicity. 

For interactions with bidirectional information flow we are able to define a criterion for a layered structure of systems: we name a bidirectional interaction ''horizontal'' if all interacting systems behave the same with respect to state, determinism and synchronicity and we name it ''vertical'' --- providing a semantic direction --- if there is a behavioral asymmetry between the interacting systems with respect to these properties. 

It is shown that horizontal interactions are essentially stateful, asynchronous and nondeterministic and are described by protocols. Vertical interactions are essentially top-down-usage, described by object models or operations, and bottom-up-observation, described by anonymous events.

The reference model thereby helps us to understand the significant relationships that are created between interacting discrete systems by their interactions and guides us on how to talk about discrete system interoperability. 

To show its conceptual power, we apply the reference model to assess several other architectural models, communication technologies and so called software design or architectural styles like SOA and REST.
\end{abstract}

%
\section{Introduction}
%
The enormous growth of the internet has been mainly due to semantically agnostic information transport protocols like HTTP, FTP, SMTP, etc., creating a huge network of interconnected devices. Meanwhile the devices overtake more and more the processing of the exchanged information, thus bringing the issue of ''semantic'' interoperability to the fore. To ''interoperate'' means that the information processing in all involved systems is accomplished in a way that, by interaction, a specified purpose is fullfilled \cite{IEC_60050}. 

In fact, a plethora of different technologies has evolved, each with the promise to solve some aspects of the interoperability puzzle. However, different underlying models, in conjunction with different terminologies, hinder mutual understanding and drive the costs for successful device integration. 

Insofar, an easily comprehensible reference model of interaction semantics to guide the discouse on interoperabilty is highly desirable. A reference model in this sense is a ''conceptual framework for understanding significant relationships among the entities of networking systems, based on a small number of unifying concepts'' \cite{OASIS2006_SOA_RM_1}. 

The contribution of this paper is to provide such a reference model of interaction semantics which is based on a simple classification of system interactions. It can be viewed as an extension of the OSI-model \cite{ISO_OSI_1994} based on sound semantic principles. 

We use the term ''semantic'' as a synonym for ''with respect to processing'' to emphasize the distinction between information transport and information processing. In this sense, we say that information gets transported and the meaning of the information, its significance, is attributed by proccessing.

The structure of the article is as follows. In the second section we lay out the formal definitions of the basic notions which we need to build our reference model. In the third section we introduce a simple classifications of system interactions based on information flow and processing. In the forth section we apply our reference model to a selection of related approaches. And in the last section, we summarize a couple of direct consequences of our reference model.

\section{The model of system interaction} \label{s_model}

In this section we provide formal definitions for the basic notions that our reference model is based upon. Due to the scope of this article, we will use them only to make the meaning of our reference model more explicit and not to derive more intricate relationships between them (For more details, see e.g. \cite{Reich2016_systems}). 

\begin{figure}[htbp]
  \begin{center}
    \includegraphics[width=5cm]{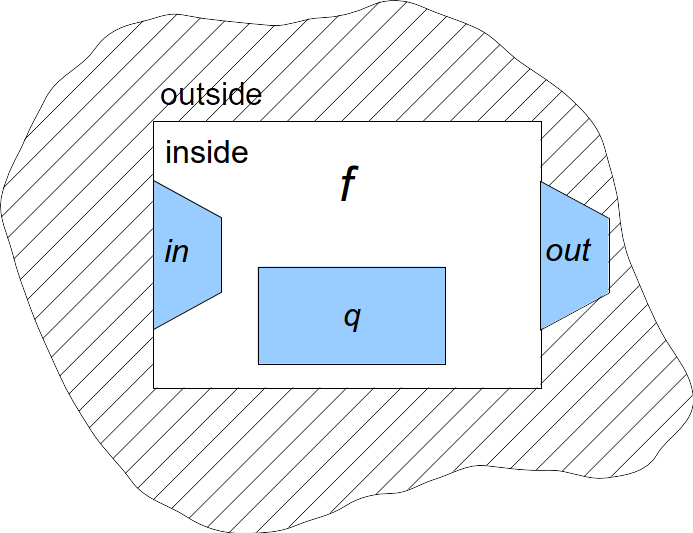}
 \end{center}
\caption[]{A symbolic representation of a system with its three signals $(in, out, q):T\rightarrow Q \times I^{\epsilon} \times O^{\epsilon}$ and its system function $f$, separating the inside from the outside.} \label{fig_system}
\end{figure}

The basic entities of our reference model are discrete systems. There seems to be a consensus (e.g. \cite{IEC_60050}) that a system separates an inside from the rest of the world, the environment (see Fig. \ref{fig_system}). We define what we call a discrete multi-input system to consist of a set of discrete time values $T$ together with a time function $succ$ which always provides the next point in time; three signals, an input signal $in$, an output signal $out$ and an internal signal $q$, each mapping time onto the respective alphabet $I$, $O$, and $Q$; and a system function $f$ mapping input and internal signal values at time $t$ onto output and internal signal values at time $t'=succ(t)$. The system is called a multi-input system because it could well be that the input characters are vectors, where some components for some characters remain empty. With the convention that $\epsilon$ is the empty character and for any alphabet $A$ we write $A^\epsilon = A \cup \{\epsilon\}$, we therefore have the following formal definition:  

\begin{definition} \label{def_system_multi-input}
A discrete system ${\mathcal{S}}$ is defined as ${\mathcal{S}} = (T, succ, Q, I, O, q,$ $ in, out, f)$. $Q$, $I$ and $O$ are alphabets, whereas only $Q$ has to be non-empty. The signals $(q, in, out):T\rightarrow Q \times I^{\epsilon} \times O^{\epsilon}$ form a discrete system for the time step $(t, t'=succ(t))$ if the partial function $f:Q \times I^\epsilon \rightarrow Q \times O^\epsilon$ with $f =(f^{int}, f^{ext})$ maps as following 
\[{q(t') \choose out(t')} = {f^{int}(q(t), i(t)) \choose f^{ext}(q(t), i(t))}\,.\]
If the input signal contains two or more components, we also call it a ''multi-input system'' (MIS).
\end{definition}

We can describe the behavior of a discrete system with an I/O-transition system. Thereby we get rid of the explicit time dependency, exploiting its stepwise character. Additionally we are able to look at the behavior of a discrete system in the sense of a projection, which becomes essential for protocols.  

\begin{definition} \label{def_transition_system}
An I/O-transition system ${\mathcal A}$ is defined as ${\mathcal A} = (Q, I, O, q_0, \Delta)_{\cal A}$. 
$Q_{\mathcal A}$, $I_{\mathcal A}$ and $O_{\mathcal A}$ are alphabets, whereas only $Q$ has to be non-empty. $q_0\in Q$ is the initial value and $\Delta_{\mathcal A} \subseteq I^\epsilon \times O^\epsilon \times Q \times Q$ is the transition relation. 

Such an ${\mathcal A}$ describes the behavior of a discrete system  ${\mathcal{S}}$ in the sense of a projection, if $Q_{\cal S} \subseteq Q_{\cal A}$, $I_{\cal S} \subseteq I_{\cal A}$, $O_{\cal S} \subseteq O_{\cal A}$, $q(0) = q_0$ and a projection function\footnote{A projection function $\pi$ fulfils the  property $\pi = \pi \circ \pi$.} $\pi = (\pi_Q, \pi_I, \pi_O): Q_{\cal S} \times I_{\cal S}^{\epsilon}\times O_{\cal S}^{\epsilon} \rightarrow Q_{\cal A} \times I_{\cal A}^{\epsilon}\times O_{\cal A}^{\epsilon}$ exists and $\Delta_{\cal A}$ is the smallest possible set, such that for all times $(t, t'=succ(t)) \in T_{\cal S} \times T_{\cal S}$, and for all possible signal runs it holds $(\pi_Q(q(t)), \pi_Q(q(t')), \pi_I(in(t)), \pi_O(out(t'))) \in \Delta_{\cal A}$.
\end{definition}   

We can say that our model of information processing treats information as characters of an alphabet and presupposes a processing context in the form of an I/O-transition relation.

\begin{figure}[htbp]
  \begin{center}
    \includegraphics[width=5cm]{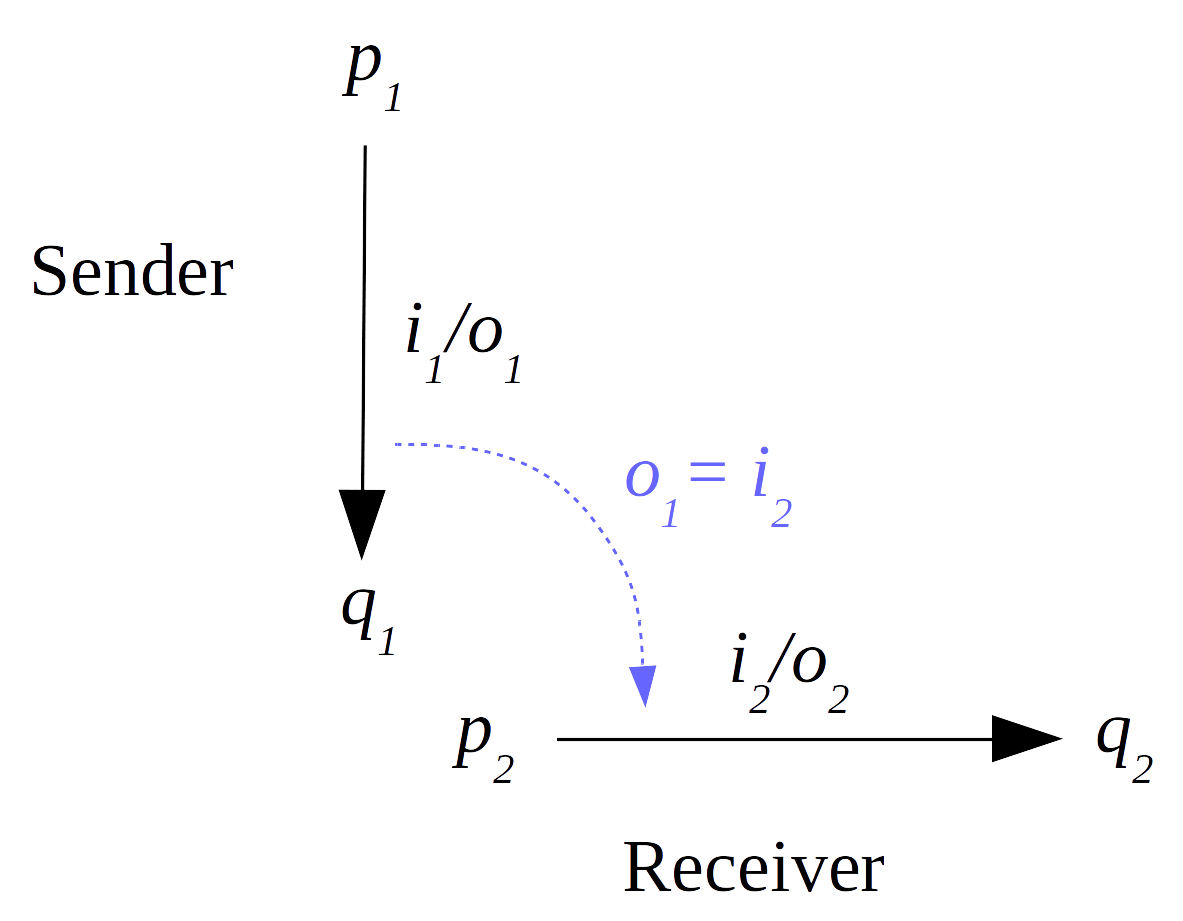}
 \end{center}
\caption[]{Interaction between two systems where the output character of a sender system is used as an input character of the receiving system.} \label{fig_coupling_by_interaction}
\end{figure}

As is illustrated in Fig. \ref{fig_coupling_by_interaction}, interaction between discrete systems means that they share a common signal: The output signal of a ''sender'' system is identical with the input signal of a ''receiver'' system. Thereby the output value of a transition of a ''sender'' system serves as the input value of a transition of a ''receiver'' system. In other words, interaction means that information is transported and the execution schema of the product automaton has to be modified such that a 'transported' character has to be processed next.

The interaction mechanism is thereby formally based on identically named input and output characters of otherwise anonymous transitions\footnote{There are many interaction models based on transition systems with named transitions, where the coupling of different systems is achieved by identically (or complementary) named transitions, e.g. \cite{Hoare1985,Milner1992}.}. 

For our classification of system interaction in the next section, we define the following behaviors for sender and receiver systems. 
\begin{itemize}
\item A sending system behaves synchronously if the completion of the receivers transition is a necessary and sufficient requirement for the sending system's next transition. Otherwise it behaves asynchronously.
\item A receiving system behaves deterministically if its transition relation represents a transition function. Otherwise it behaves nondeterministically
\item A receiving system behaves statefully if its set of internal state values $Q$ has more than one element. Otherwise it behaves stateless.
\end{itemize}
Please note that in our definition, synchronicity is a property of the sending system. That is, whether a sender waits or not (behaves synchronously or asynchronously) is not detectable by the receiver, processing any received character. 

Whether an interaction is deterministic or nondeterministic influences the kind of goal of the interaction. In the first case, the goal of the interaction is a superordinated function, leading to system composition \cite{Reich2016_systems}.
In the second case, without creation of a superordinated system function, the need to distinguish desirable from non-desirable behavior necessitates an additional acceptance component, extending the I/O-transition systems to become I/O-automata or transducers, leading to protocols. 

\section{Classification of system interaction}
The essential idea of the reference model is that system interactions can be classified along the two relevant aspects of interoperability: information transport and information processing \cite{Reich2015_kvsi}. Information transport can be either unidirectional or bidirectional. Information processing is further subclassified along the three sub-dimensions state, determinism and synchronicity (see section \ref{s_model}).

We chose these three subdimensions of information processing because of their direct influence on the form of the appropriate interface, where we understand an ''interface'' as the shared boundary across which two or more systems exchange information. For example: there are no return parameters in the asynchronous case. In the deterministic case, it is possible to represent the intended input-output-relation of a system by an operation, mapping state values onto state values. With statefulness and determinism the desired functionality can be described object oriented and last but not least, we describe stateful, nondeterministic and asynchronous interactions as protocols. 

This classification is complete in the sense that every interaction that can be described with the formalism of section \ref{s_model} can be classified accordingly.

\subsection{Interaction with unidirectional information flow}
As backward communication is irrelevant for unidirectional interfaces, we can disregard any synchronicity. The two most important classes are:\\[-0.3cm]

\noindent
{\bf Deterministic:} We name a sequence of systems a ''pipe'', where an overall computational function is computed in a number of successive steps on a ''data flow'', where the input of each pipe component is the output of the predecessor component (except for the first one). Thereby pipes provide the means for sequential and parallel system coupling in the sense of \cite{Reich2016_systems}. To be complete, a pipe mechanism must be able to fork and join pipes. \\[-0.3cm]

\noindent
{\bf Nondeterministic:} We name an interaction between a sender system and a receiver system an ''observation'' if the sender system makes no assumptions on the determinism and statefulness of the receiver system.   

\subsection{Interaction with bidirectional information flow}
As it is bidirectional, the flow of information as such does not determine any direction of the interaction relation any longer, but the direction is determined by the way, the information is processed in all interacting systems - and therefore is a semantic property. We distinguish two main bidirectional interaction classes:\\[-0.3cm]

\noindent
{\bf Horizontal interaction:} 
All interacting components behave the same with respect to the three semantic sub-dimensions, i.e. there is a behavioral symmetry. Only the combination stateful, nondeterministic and asynchronous behavior makes sense. Mutual determinism results in a deadlock where each system waits for some input. Nondeterminism and statelessness implies randomness. And mutual synchronous behavior makes only sense in the calculation of recursive functionality. 

Horizontal interactions are described by bi- or multilateral interfaces of protocol roles. This multilaterality manifests itself by the fact that the knowledge of all roles of a protocol is necessary to guarantee important properties of this form of interaction.\\[-0.3cm]

\noindent
{\bf Vertical interaction:}
 the interacting components behave differently with respect to the three semantic subdimension. The resulting asymmetric setting creates a semantic direction of the interaction. The ''lower'' component behaves deterministically and can therefore be described by a function call (with exceptions). We say that it does not make any assumptions with respect to the behavior of the ''upper'' component and can therefore provide information upwards only by an event-mechanism that is similar to the observation in the case of unidirectional information flow. Actually, within such an interface, only the lower component is described with its functionality and its events --- why we call these interfaces ''unilateral''.

\subsection{Components and their hierarchies}

\noindent
{\bf Components:}
An important concept that is touched by our reference model is that of components. Components are supposed to be building blocks which easily fit together (e.g. \cite{HeinemanCouncill2001,Szyperski2002}). 

Our reference model suggests to use both, the characteristics of a system's I/O relation as well as its composition behavior, to define the component concept. 
Two systems might comprise the same I/O-relation but may differ in their composition behavior like operations versus pipes. Both are deterministic (according to our definitions), but operations provide their output back to the component where they received their input from, while pipe components provide their output to the ''next'' component of the pipe. Additionally, protocol based nondeterministic interactions do not lead to supersystem formation. 

Bidirectional interactions in a recursive loop or while sense, creating complex recursive functionality, do not follow any simple composition scheme and should be avoided on the level of components. Thereby, from an software engineering point of view, components also mark a systematic border of design complexity, where any functionality that is created by general recursion moves into the component’s innards. \\[-0.3cm]

\noindent
{\bf Layers:}
Our rather simple classification allows the definition of a layering in a component based system and thereby relates interoperability to these different layers. 
Components that interact vertically belong to different layers, components that interact horizontally belong to the same layer. Observed components belong to lower layers and all components of a pipe belong to the same layer. 

\begin{figure}[htbp]
  \begin{center}
    \includegraphics[width=9cm]{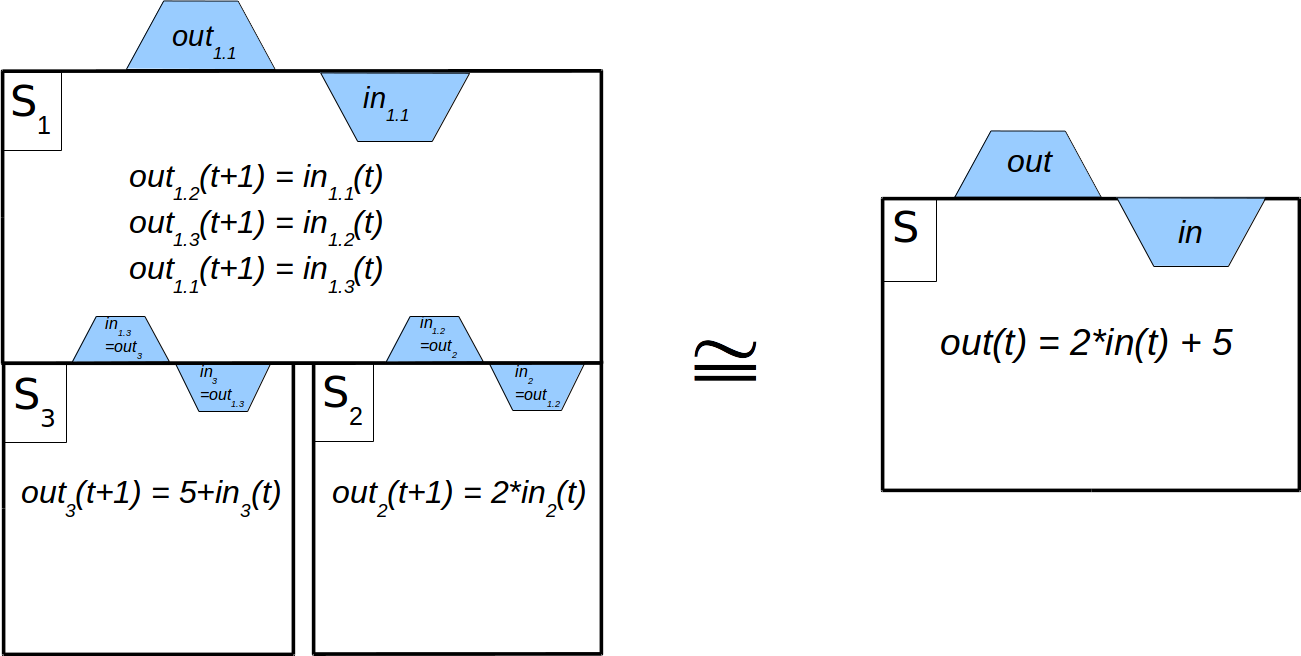}
 \end{center}
\caption[]{Three systems ${\mathcal S}_1$, ${\mathcal S}_2$, and ${\mathcal S}_3$ compose to a supersystem ${\mathcal S}$ with the overall (super)system function $f_{\mathcal S}(x) = 2x+5$.} \label{fig_example_composed_system}
\end{figure}

Please note, that one has to be precise to what kind of hierarchical relation one refers. To illustrate the problem, we provide a simple example. Fig \ref{fig_example_composed_system} shows a simple system composition where three systems ${\mathcal S}_1$, ${\mathcal S}_2$, and ${\mathcal S}_3$ compose to a supersystem ${\mathcal S}$ with system function $f_{\mathcal S}(x) = 2x+5$. System ${\mathcal S}_2$ contributes its system function $f_{{\mathcal S}_2}(x) = 2x$, ${\mathcal S}_3$ contributes $f_{{\mathcal S}_3}(x) = x+5$, and System ${\mathcal S}_1$ is a multi-input system which mostly coordinates system ${\mathcal S}_2$ and ${\mathcal S}_3$ in a non-trivial recursive manner. 
As we can easily see, there is no interaction between the subsystems and their supersystem, but instead, the supersystem is created by the deterministic interactions of the subsystems.   

\begin{figure}[htbp]
  \begin{center}
    \includegraphics[width=9cm]{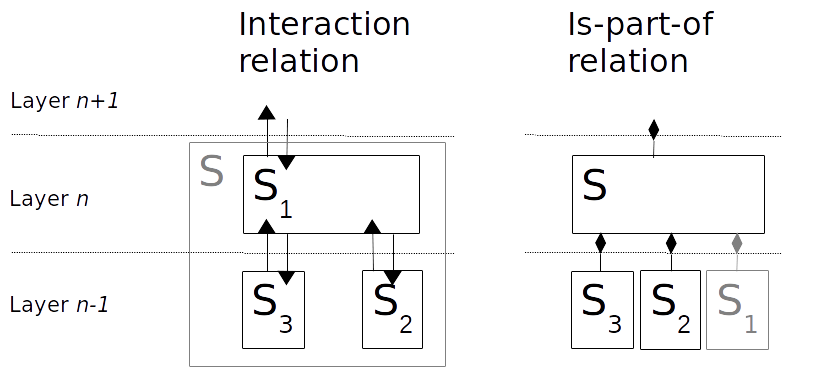}
 \end{center}
\caption[]{Due to their interactions, the systems of Fig. \ref{fig_example_composed_system} can be ordered in two different ways. 
On the left side, they are ordered according to their interaction relation. The arrows represent the information flow. The supersystem ${\mathcal S}$ is only sketched in light gray to show that it does actually span several layers in this hierarchy. Or they can be ordered according to their ''is-part-of''-relation, shown on the right side. Now it is the supersystem ${\mathcal S}$ that is super ordinated. The relation is represented by a filled diamond and a solid line. There is no information flow between the layers.}  \label{fig_example_hierarchy}
\end{figure}

According to our classification, system ${\mathcal S}_1$ interacts hierarchically with both ${\mathcal S}_2$ and ${\mathcal S}_3$ as it determines their transitions by its output and not vice versa. Thereby it is justified to say that because of their interaction, system ${\mathcal S}_1$ belongs to a higher layer as both, systems ${\mathcal S}_2$ and ${\mathcal S}_3$. The supersystem ${\mathcal S}$ is not mentioned at all in this description. This is shown in the left side of Fig. \ref{fig_example_hierarchy}. 

However, we could also define a hierarchical relation in the sense of ''is-part-of'' where both subsystems, ${\mathcal S}_2$ and ${\mathcal S}_3$, are part of the supersystem ${\mathcal S}$, which is shown in the right side of Fig. \ref{fig_example_hierarchy}. Now, system  ${\mathcal S}_1$ is not mentioned any longer. 

This latter hierarchy is used in imperative programs and the object oriented world with their method-construct. A method represents a function which --- if not elementary --- depends on other methods. Then, the formal test for the claim, that a component can be put into a certain layer is to provide a unilateral interface with generic events and operations and to assure that the component itself uses only operations and reacts to generic events of components of lower layers.  

With the ''is-part-of''-relation, the higher the layer, the more abstract --- or the less technical --- is the level of information processing.\\[-0.3cm]

\noindent
{\bf Remote operations}
Remote operations take advantage of the fact that a ''remote'' operation can be partitioned into a sequence of concatenated operations of serialization, deserialization, transport and local processing.

In the case of a remotely used object, the communcation components of both sides become a ''communciation layer'' that can indeed be hidden behind the remote operation's signature --- adding, however, remote exceptions. Thus, despite the similarity between a local and a remote object usage, the imponderables of information transport usually introduces nondeterminism. Due to their (much) higher unreliablity, remote operations should not be used to change remote state. \\[-0.3cm]

\noindent
{\bf Protocols}
A protocol (see e.g. \cite{Holzmann1991} for an overview) is best understood as a signaling game in the sense of P. Grice \cite{Grice1989_Studies} where all participants are only described as a ''role'' in the sense of a projection, giving each other mutual hints and rely on the fact that their sent information will be appropriately processed. In our model, each participating system can be described in its participating role by an automaton according to definition \ref{def_transition_system}. A protocol has to be ''wellformed'' in the sense that for all sent characters there has to be a receiving transition, ''complete'' such that there are no other inputs, ''interuptable'' in the sense that it does not have infinite chains of interactions, and ''consistent'' such that it should be possible to meet the acceptance condition from each reachable state. 

As P. Grice pointed out we can distinguish between an assumed receiver semantics of the sender and the actual semantics of the receiver. As both relate to the processing of the receiver, they can easily be compared. 

We think that extending the interaction description of protocols by ''decisions'' as an additional internal input alphabet to fill in the non-determinism leads to the game-notion \cite{Reich2009}. 

%
\section{Application of the reference model}
%

\subsection{The open systems interconnection basic reference model}
The Open Systems Interconnection (OSI) basic reference model \cite{ISO_OSI_1994} was very influential, as it established the notion of a layered software architecture. 

However, the OSI model assumption, that ''OSI is concerned with the exchange of information between open systems (and not the internal functioning of each individual real open system).''  turned out to be inconsistent. One cannot refrain from saying anything about the structure of information processing and at the same time making claims about its inner structure, like layering. Also, the OSI model was not sufficiently precise with respect to the kind of hierarchy.

Thus, our reference model provides a formal justification of the intuition of the OSI-model to view software applications as being layered. Thereby, our model also explains, why the OSI-model found its way into reality only up to its 4-th layer, as the mangement of a ''session'' state cannot be attributed to a dedicated layer in the general case.  Only in the case of vertical interaction, the interaction related state can be encapsulated into a state of an intermediate ''session layer''. In the case of horizontal interaction, the interaction related state genuinely belongs to the components of the same semantic layer that mutually interact.

\subsection{Interaction pattern based approaches}
There exist quite a few approaches that present themselves as ''pattern based'' and often claim to be motivated by the speech act theory \cite{Austin1962_Howto}. Examples are the UN/CEFACT modelling methodology \cite{UMM2003UserGuide} or RosettaNet \cite{RNIF20} using ''transaction patterns''. A similar concept had been the ''message exchange pattern'' \cite{WSDL20} of the service oriented architecture (SOA).

The application of our reference model already fails at the very beginning, as these approaches do not relate the interaction semantics sufficiently to the transformational behavior --- the processing capabilities --- of the interacting systems. 
As was shown in \cite{Reich2008}, without referring to the transition relation of sender and/or receiver, the semantics (=processing) of any ''interaction pattern'' is illdefined. 

An example for a total lack of reference to the transformational behavior was SOA. The idea of a SOA was introduced by Roy W. Schulte and Yefim V. Natis of Gartner in 1996 \cite{SchulteNatis1996} and it was endorsed by virtually all large IT companies like Microsoft, IBM, SUN, Oracle, Adobe, SAP, etc. Within the OASIS SOA reference model \cite{OASIS2006_SOA_RM_1}, a service is defined vaguely as a ''Mechanism to enable access to one or more capabilities where the access is provided by a prescribed interface and is exercised consistent with constraints and policies as specified by the service description.''
At no place within SOA the service definition refers to some transformational behavior, making any distinction between deterministic vs. nondeterministic interaction impossible. But, from a syntactic point of view, SOA ties oneself down to (remote) objects with methods. Thus, well defined services specified by a WSDL-specification can only represent accessible functionality --- and not multilateral interfaces of protocol participants.

\subsection{Representational State Transfer (REST)}
REST \cite{Fielding2000} can be viewed as the attempt to transfer the principles of stateless communication together with semantic agnostics - both principles of the HTTP-protocol - onto the interactions of networking applications. Currently it is often positioned as a simpler variant of SOA.

A REST-service is supposed to adhere to the principle of addressability, that each resource has to have a unique URI, and statelessness, that each REST-message is supposed to contain all the information that is necessary for the processing which it initiates.
Sometimes idem potency (e.g. \cite{Pautasso2014}) is mentioned, that the called transport methods are supposed to have an identical effect, no matter when they are called.

These ''principles'' are in direct contradiction to the proposed interaction model where horizontal interactions usually are stateful and the exchanged information is usually not processed in an idempotent way.  
From the perspective of the proposed interaction model, REST is a methodological chimera with parts from the object as well as the protocol world. On the one side, it only specifies a letter-box mechanism, which gives some of the letter life cycle functionality to the sender. It does not require to specify the transformation behavior of a REST-service or the relation between different REST-services. But on the other side, it requires that all resources have to be published to the public, offering a lot of otherwise private information.

\subsection{Evaluating interaction supporting technologies}
One strength of our reference model is to allow a simple assessment of whether a given interaction oriented technology supports a certain interaction class.  First, it separates these technologies into information transport versus information processing categories, depending on whether they relate to the content of the transported information or not. 

Then it makes clear that content-oriented technologies supporting only remote object models like DCOM or OPC-UA are of little value when it comes to implement protocol based interactions. 

It also explains, why it is inappropriate to use technologies primarily supporting unidirectional observation like publish-subscribe for bidirectional horizontal interactions, as is the case in many ''service-bus''-models. In a unidirectional observation the sender does not make any assumptions about the processing of the receiver. Thus, the created events either have a strongly standardized format, like instance X of type Y changed its state from Z1 to Z2 - making mutual understanding impossible. Or the created events become arbitrarily broad to contain all the information that is possibly available, just in case, the (unknown) receiver might find it necessary.
 
\subsection{Examples of inadequate usage of the layer concept in models of interoperability}
Often, architectural models provide a layer concept, without stating any verifiable ordering criterion. As a result they mix categories and provide at most a purely intuitive level of understanding, too little in case some engineer wants to learn something for her system construction. 

One example is the ''Level of Conceptual Interoperability Model (LCIM)“ \cite{TolkTurnitsaDialloWinters2006}, which consists of the 7 alleged layers: no [interoperability], technical, syntactic, semantic [defined not in our sense], pragmatic, dynamic, and conceptual interoperability. Obviously, it is not interaction which can account for this hierarchy, but what else? Even for the technical information transport e.g. by the internet protocol, a certain structure (=syntax) of the transported information is necessary. It is unclear how to separate semantic from pragmatic aspects. For example, how can the meaning of a bank tranfer be described without refering to some action a bank is supposed to take? Etc.   

Another example is the architecture axis of the three-dimensional Reference Architecture Industry 4.0 (RAMI4.0) \cite{DIN_SPEC_91345:2016-04}. It consists of the 6 alledged layers: asset, integration, communication, information, functional and business. Again it is not interaction that accounts for this hierarchy, But what else? 

A third example is the IIC Connectivity Framework \cite{IIC2018_Connectivity_Framwork}. It builds upon the LCIM. The transport layer is supposed to achieve technical interoperability and a framework layer is supposed to achieve syntactical interoperability in the sense of providing all means to exchange structured data. However, determining the type of the exchanged data also determines whether it is to be understood as an operation, as a business document or a generic event --- requiring an agreement on the interaction class beforehand. As a result, the authors neglect to discuss how to support the important class of horizontal interactions.     

In effect, most of these so called ''layers'' are in fact aspects, that is, something that can be described from a certain point of view in the sense of a projection. Thus, we have data aspects, communication aspects, functional aspects, etc. But all these aspects overlap and do not constitute a hierarchical division of the thing of interest.

%
\section{Conclusion}
%
We proposed a reference model for interaction semantics. It is supposed to guide the discourse on interoperability, what the real challenges are, how to achieve them, etc. for example in standardization efforts. Based on a proposal of one the authors, it has already been adopted by the German VDI/VDE Fachausschuss GMA 7.20 for their VDI-guideline 2193 ''Semantik und Interaktion von I4.0-Komponenten'' which will be published in 2019 and it has become the base of a current BITKOM effort to provide guidance for the industry on inter-domain issues of interoperability.

Our model is conceptually sparse as it uses very little ad hoc assumptions. It is based on a sound system model, fully compatible with the model of information transport and processing. And it is expressive as it entails a whole series of important consequences:

\begin{itemize}
\item It shows the importance of a precise and verifiable order criterion for a well defined layered structure of systems. 
\item It is very important to distinguish between interaction scenarios with unidirectional and bidirectional information flow. Especially, an interaction with bidirectional information flow is not just the superposition of two otherwise independet single-flow scenarios. Adding an information flow ''backwards'' in a formely single-flow scenario changes the game of interoperability fundamentally.  
\item Terms like ''message based integration'' are ill defined, as they suggest that the processing context of the transported information can be ignored. Looking only at the act of information transport, one cannot say --- per definition --- anything about the processing of the information, for example, whether the processing of the information is stateful, synchronous or deterministic.
\item Any attempt to propose some inner structure of information processing systems without reference to the structure of information processing itself is inconsistent.
\item Horizontal interactions are nondeterministic, asynchronous and stateful. 
\item Technologies providing only access to remote functionality in the sense of object models, are of little help to implement horizontal interactions.
\item To easily structure applications into layers, an internal event mechanism is needed to avoid the use of operation calls to provide information for higher level processing. Only without circular functional dependencies does an operation call formally indicate a descent into a lower software layer. 
\item We have to distinguish between transformational and compositional behavior, leading to the difference between systems and components. 

\end{itemize}

\bibliographystyle{IEEEtran.bst}
\bibliography{philosophy,informatics,soziologie}

\end{document}